\documentclass[aps,prl,twocolumn,groupedaddress, showpacs, super,comma]{revtex4}

  \usepackage{graphicx}
  \usepackage{color}

\begin{document}

\title{ Multi-quanta Abrikosov vortices as the micro spin batteries }


\author{Zygmunt Bak}
\email[]{z.bak@ujd.edu.pl}
\homepage[]{}
\affiliation{Institute of Physics, Jan Dlugosz University of Czestochowa,
42-200 Czestochowa, al. Armii Krajowej 13/15, Poland.}


\date{\today}

\begin{abstract}

The aim of our paper is to study the multi-quanta Abrikosov vortices injected into a superconductor layer by the twisted light impulses. We predict that the condensate circulating around the core of a multi-quanta vortex may differ in nature from that of the bulk superconductor. Special attention is paid to possibility of appearance of the quasi angular momentum of Cooper pairs and its compensation via their recombination to the fermion quartets. We show potential applications of the multi-quanta vortices as massive memories as well as spin, angular momentum and energy batteries. Finally, we predict the emergence of half-quantum multi-quanta Josephson vortices under certain conditions.

\end{abstract}


\maketitle


\section{Introduction}

In the type II-superconductors, beyond some critical limit of the external magnetic field, arise supercurrent vortices circulating   around normal cores \cite{abrikosov}. The supercurrents forming the Abrikosov vortex induce magnetic field with the total magnetic flux $\Phi _o = h/2e$ /fluxon/. As the rule, spontaneously formed Abrikosov vortices are single valued i.e. each carries a single fluxon and form regular /Abrikosv/ lattices. Nevertheless, along with the  experimental verification of early predictions of a ground state consisting of a lattice of singly quantized vortices, many reports on their nucleation have been made (see \cite{barnett} and references therein). In laminar system vortices are strongly influenced by the sample geometry, the lateral confinement may even force them to merge into  multiquanta vortices otherwise unstable in bulk geometry \cite{kanda, kadowaki}.

With the intention of technological applications we cannot rely on the spontaneously formed  multi-quanta vortices. The simplest way to create them in controlled manner is the excitation by the vortex light beams. Conventional optics involves photons that carry only linear momentum, however, light can carry both linear and orbital angular momentum /OAM/. The latter have a spin component associated with polarization and an orbital part associated with the spatial structure \cite{bliokh}. The light beams with nonzero angular momentum are referred as the {\it twisted light} or {\it vortex beams}. Useful feature of optical beams carrying orbital angular momentum is that it can be easily manipulated and transferred. This opens new horizons across various fields, much like the use of additional degrees of freedom in other types of excitations, such as electrons \cite{szczesniak2020}. Notably, new applications range from mechanical micromanipulation \cite{padgett3} and imaging science \cite{maurer4} to communication technologies \cite{wang8}. Beyond optical wavelengths, OAM now plays a major role in electron \cite{uchida9, karimi12} engineering. In biophysics and micromechanics, OAM induced torque has been proposed to be utilizable in driving molecular motors and micro-machines \cite{brachmann}. It has been also shown that with the vortex light beams we can excite twisted magnons, excitons, polarons or electron wave-packets.

The aim of our research is to study the interaction of vortex beams with magnetic and Abrikosov vortices in solid laminar systems. We focus our research on theoretical study of various AOM excitations and its propagation in a multilayered system consisting of an isolator film sandwiched between two superconducting layers. It is interesting that for relatively thin normal interlayer we have the Josephson junction
and under some conditions a half-valued vortices /Josephson vortices/ can be formed \cite{goldo}. Adjoint half-valued Josephson vortices each carrying  the semifluxon can merge with Abriksov vortex and form multi- half-valued items. In the following, attention will be paid to the formation of multiquanta,  Abrikosov-like vortices generated either by the external magnetic field or by the injection via OAM
photons. Finally, we predict a novel form of the Hall-like effect due to the interaction of OAM beams with the multiquanta Abrikosov vortices.
 
\section{Vortex beams}

Theoretical studies of the OAM excitations have not only cognitive character. The practical benefit is that the orbital angular momentum allows OAM light/excitation to carry more bits of information per photon than spin alone, so the ability to manipulate the orbital component could open up new communication methods (e.g. in the 5G telecommunication \cite{fouda} or speed up data transfer). In addition, it is evident that controlling the interaction of spin and orbital angular momentum for the same quasiparticles may allow for novel kinds of logic operations in future optical devices.

The use of OAM current, a flow of twisted excitations, instead of the charge or spin current is considered as a possible route to overcome the limitations of the Moore's law. In the construction of faster and  more compact computing  circuits information processing is inevitably accompanied by a certain amount of heat generation. The problems with removing redundant heat makes that studies of the problem turn back to the superconducting /SC/ memories which greatly reduce the dissipation power and increase the speed of operation. Especially as maintenance costs exceed the cost of refrigeration to cryogenic temperatures \cite{golod}. Abrikosov vortices, representing spatially isolated single magnetic flux quanta in superconductors (SCs), can be inherent carriers of digital information in advanced microprocessors and memories \cite{vlasko, golod}. In some cases magnetic-flux qubits are preferable to charge qubits, holding potential to overcome some limitations such as the flux noise \cite{szczesniak2021}. Until now all concepts of Abrikosov lattice applications were limited to the monoquanta vortices.

In the following we will discuss the OAM photon induced multiquanta Abrikosov vortices that offer new possibilities of technological applications. Knowledge about the structure and dynamics of collections of vortices is of importance both to the understanding of the basic physics of superconductors and to the design of devices. The use of multiquanta Abrikosov vortices in the construction of computer memories increases its capacity. Simultaneously superconductivity allows reduction of the redundant heat generated.

The interaction of twisted light with a solid is always associated  with exchange of   both linear and angular momentum. Such a type of excitation allows creation of  a variety of quasiparticles which carry the angular momentum. When the electron, exciton etc. possesses orbital momentum the Schroedinger equation  written in the cylindrical coordinate system ($\rho$, $\theta$,  z) gives the solutions that are no longer plane waves,
\begin{equation}
\label{e1}
\Psi _p (\vec{r}, t) \propto e^{i(\vec{p} \cdot \vec{r} -E(p)\cdot t)/h} ,
\end{equation}
with $\vec{r}$ being  the position operator, but instead are given by the monochromatic
twisted (vortex)  Bessel beams \cite{bliokh}.
\begin{equation}
\label{e2}
\Psi _{p_\rho, p_z}\ = \ |p,l> \ =\alpha\;   J_{|l|} (p_\rho \cdot
 \rho/h) e^{i(p_z\cdot z -E\cdot t)/h } e^{il\theta}  ,
\end{equation}
or alternatively   the Laguere-Bessel /LG/ beams
\begin{equation}
\label{e3}
\begin{array}{lll}
\Psi ^{LG}_{l,n} \propto & \left( \frac{r}{w(z)} \right) ^{|l|}  L_n^{|l|}\left( \frac{2r^2}{w^2(z)} \right) \times \\
& \times exp\left( -\frac{r^2}{w(z)}^2 + ik\frac{r^2}{2R(z)} \right)   e^{i(l\phi +kz)} .
\end{array}
\end{equation}
This wave-function is the product of three types of functions: an plane wave with momentum $p_z$ in the z-direction, a radial component written as a Bessel function of the first kind $J_{|l|}$, where the $p_\rho $ is the linear momentum in the radial direction, and finally an azimuthal component written as $exp(il\theta)$ where $l$ (i.e. $m_z$) represents the magnetic quantum number related to the angular momentum $ L_z$ in the z-direction. Twisted light interacts with matter differently than spin does. If a particle sitting off-axis in a light beam absorbs a photon with orbital angular momentum, it responds by circulating around the beam, not by spinning
on its axis.

The $l>0 $ beams show  quantum vortex $exp(il\phi)$, spiral phase structure, nonzero azimuthal probability current, and the probability density vanishing on the axis: $|\Psi |^2|_{r=0}$=0. Gaussian beams or wave-packets are often implied in quantum models of free electrons, because they do not contain any intrinsic structures and degrees of freedom. In contrast to that, higher-order modes with (l,n) $\neq $ (0,0) exhibit a variety of structures related to the different (l,n) or Bessel beams with different $l$, constitute a complete set of orthogonal the monoenergetic modes for the free-space Schroedinger equation.

For such  beams the wave-front has a helical spatial structure  around beam  axis.
This refers not only to the  optical waves but to any other, induced by the OAM beams,  solid state excitations. For example twisted electron beams can be produced at scanning electron microscopes with energies up to keV and orbital momentum as high $m\hbar = 1000\hbar$ (see \cite{karlovets} and references therein). However, with ultrashort pulses we are interested in, there is a fundamental restriction to the topological charge of the vortex, and hence upper bound to the OAM, carried by a pulse. \cite{porras} Generally, from the monochromatic waves the localized  wave packets (with the same $m$ in the bundle) can be formed to produce required properties useful in technological applications \cite{schultze}. Recently many forms of two-dimensional, structured light beams like Weber beams, Mathieu beams, vector beams, and beams with arbitrary transverse shapes have been found \cite{drake}. In the following we limit consideration to the Bessel (\ref{e2}) and LG (\ref{e3}) beams.

The fact that OAM beams can transfer both spin and angular momentum to the vortex opens the question of the nature of superconductivity that forms circulating supercurrents so one would expect appearance of unconventional superconductivity associated with the multi-quanta vortices. As we know the $s$-state Cooper pair states easily develop in an isotropic medium, however, in laminar systems when layer thickness is less than coherence length s-pairing is not favoured.  The bulk condensate form the Cooper pairs which are not coherent with the vortex supercurrents, this means that both condensates, bulk and the vortex, 
are completely decoupled. The latter forms a droplet that rotates with some angular velocity $\omega $, thus both condensates can be of different nature. Moreover, some experiments suggest  spatial segregation of pairs formed of different electronic states \cite{bianconi}, \cite{kagan}.
Stronger magnetic field within multiqanta-vortices, as well-as a laminar structure, should favour non-s superconductivity.
This can be e.g. the triplet pairs or even OAM endowed electron pairs. 
Additionally, the hallmark of triplet superconductivity are the host spontaneous currents at the sample surface \cite{curran}, Abrikosov vortices perfectly meet this expectation.
Available experimental techniques allow observation of vortices in nonconventional superconductors \cite{curran}. 

\section{Quasi angular momentum}

In theoretical research the focus is on intentionally generated excitations carrying the OAM. However, there can arise spontaneous formation of quasiparticles having nonzero angular or quasi angular momentum. The most known example of such item is the electron/hole Cooper pair. As we have pointed above in laminar systems when layer's thickness is less than coherence length s-pairing is not favored as the motion of electrons is squeezed by the adjacent layer potentials. This is especially interesting for superconducting systems being in the focus of our study. The  Cooper pair with electron's quasi-momenta $\hbar \vec{k}$ and $-\hbar \vec{k}$ and separated at distance of order of coherence length $\xi $ forced to in-plane motion should exhibit  quasi   angular momentum /quasi-AM/ pf order pf  $\vec{L} \propto  \vec{\xi} \times \hbar k$. As only the electrons close to the Fermi surface are active in the Cooper pair formation we can estimate $L$ as $l= \hbar \xi k_F$. The opposite helicities produce opposite quasi-AM and respective magnetic moments. Quasi momenta of Cooper pairs are quantized and at temperature of condensation all pairs should exhibit the same  quasi-momentum quantum number. At the onset of superconductivity the quasi momenta of Cooper (hole) pairs in the bulk systems are compensated \cite{brusov, bramati}. However, when there is a single Cooper pair tunneling across a magnetic barrier can feel the magnetization $M$ of the spacer and effectively exerts action of the potential.
\begin{equation}
\label{e3c}
V=V_c(r)+V_s(r)\;\vec {M}\cdot \vec{L} ,
\end{equation}
with $ \vec{M}$ and $\vec{L}$ being the barrier magnetization and Cooper pair quasi angular momentum. The $V_C$ is the spin independent potential while the  $V_s(r)$ represents the quasi momentum interaction with the magnetization of the spacer.
This means that e.g. Cooper pairs tunneling across magnetic barrier should feel different height and the ferromagnetic layer should filter the Cooper pairs   of different chirality. We should point out here that there can also arise the skew scattering represented by an asymmetric part of the transition probability \cite{bak-nov}. For us it is important that within the context of scattering experiments, the quasi angular momentum and the true angular momentum  play identical roles and cannot be distinguished \cite{paszkiewicz}.

At the interfaces the mechanisms of compensation of the quasi-AM valid the bulk \cite{brusov} are not effective, and another mechanism can be switched on to lower the barrier given by (\ref{e3}). The Cooper pairs with opposite orientation of the quasi-AM can undergo second pairing  when two distict pairs recombine into a four electron/hole aggregate (quasiparticle): a quartet \cite{quartets}. This effect can be experimentally observed by the  measurements of the Josephson current. In some systems the persistent current across the Josephson junction oscillates in time with basic period of $4eV/\hbar$ instead of $2eV/\hbar$, which can be interpreted as the effect of electron tunneling in quartets \cite{quartets}.

Such scenario can be easily illustrated in the layered copper oxide superconductors. The high in-layer mobility of electrons causes that quasi-AM of the copper pairs is oriented perpendicularly to the $CuO_2$ plane. Suppose that there is another Cooper pair in the adjacent $CuO_2$ layer with opposite orientation of the quasi-AM, then their dipolar momenta compensate each other. Theory of such four electron binding into
quartets that involves other mechanisms of coupling is presented in \cite{taras}. We should point out here that additional interactions can change the spectrum of quartets and thus the nature of transition to the SC state can be modified \cite{zbakprb}.

Alternative mechanism to explain compensation of quasi-AM of Cooper pairs at $T_c$ is a two fluid model. Cooper pairs of different chirality form a separate condensates with the same gap and the same spatial density. In this scenario opposite quasi-momenta are compensated similarly as the momenta of "spin up" and "spin down" electrons in the conduction bands of simple metals.

\section{interaction of twisted beams with vortices}

We focus our interest on a laminar system being a SC/no-SC/SC sandwich where SC denotes superconducting material. For extremely thin central spacer (of order of a few nanometers) this is just the Josephson junction but we assume rather mesoscopic dimension with larger thickness and lateral extension.

In the bulk type-II superconductor  under action of the perpendicular to the sample's surface magnetic field there arises formation of monoquanta (Abrikosov) vortices. This occurs above a certain critical field strength $H_{c1}$. The vortex density increases with increasing field strength. In thin films, when fabricated type II superconductors, even small magnetic fields $B_c \approx 1\ \mu T$ result in the creation of Abrikosov vortices i.e. regions of supercurrent that circulate a non-superconducting core. In laminar systems formation of the Abrikosov vortices is not limited to the type-II superconductors. It is well known that superconducting films made of a type-I material can demonstrate a type-II magnetic response, developing stable vortex configurations in a perpendicular magnetic field \cite{camacho}. We should point out here that vortex formation (orbital effect) strongly destabilizes the superconductivity, leading to a small upper critical field $Hc_2$. By contrast, when the field is applied parallel to the layers, most of the flux lines can penetrate between the SC layers, and then the orbital effect is strongly quenched. This is the main reason why Hc2 in the parallel direction is much higher than that in the perpendicular direction.

The way a superfluid acquires angular momentum is perhaps one of its most interesting properties. The creation of the Abrikosov lattice under action of the external magnetic field is not the only way. With the use of OAM photons we can inject multiquanta vortices into the mesoscopic SC layer. Coherent pumping, injecting OAM particles directly into the condensate at an energy $\hbar \omega $, is described by a source term \cite{keel}
\begin{equation}
\label{e3a}
\hbar \nabla _t \psi = F\cdot e^{i\omega t}e^{il\theta}.
\end{equation}
OAM phase twist, defined in terms of the rotation of the isophase surface of the   wave function, was shown to have physical, measurable effects in terms of the production of new defects \cite{foresti}. The focused incident OAM light impulse of sufficient intensity can transfer part of its energy and ngular momentum to the superfluid forming a local phase defect. The orbital angular momentum may be removed from the beam mode and converted into currents within a superfluid. As we are focused on the laminar systems the injected vortices can be the multiquanta ones. In this way there arises alternative method for Abrikosov vortices creation in one component superfluid. Recently such a direct transfer of light's OAM onto a non-resonantly excited polariton superfluid has been observed \cite{kwon}. It is possible that impact of a focused OAM photon can change the topological charge of the preexisting vortex. As the multiquatized vortex in a BEC can carry any integer number of circulation quanta the
storage efficiency of cryogenic memory in this dense-coding of information is possible. This resembles the properties of the OAM signals which also enable infinite communication channels for both classical and quantum communications. Effectively application of SC active layers as the base of cryogenic memory opens a possibility to write all of the information carried by a multiquantum OAM wave packet in a one multiquantum Abrikosov vortex.

Suppose we apply a magnetic field perpendicular to the surface, above $H_{c1}$ a net of multiquanta Abrikosov vortices is produced. If in the upper layer an Abrikosov vortex is generated its magnetic field penetrates the second layer to a usual depth $\lambda$. The screening current crosses the (normal) interlayer and induces superfluid of the other one. Provided that the thickness of the second layer is less then $\lambda $  the twin vortex in the lower SC layer is created. In this way the geometry causes that there arise a coaxial twin vortices in both SC layers (Fig. 1).

\begin{figure}[ht]
\includegraphics[height=3.5cm, width= 6.5cm]{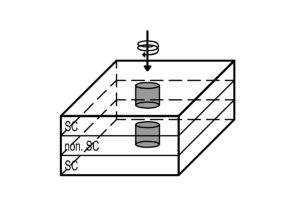}
\caption{Creation of a twin-vortices by the OAM photon impact \label{e51}.}
\end{figure}

In this way in systems composed of a few layers a pancake of vortices can be formed. It is interesting that for relatively thin normal layer we have Josephson juntion and, under some condition, a half-valued twin  vortices /Josephson vortice/ can be formed \cite{goldo}. Adjoint half-valued vortices, each carrying a semi-fluxon, can merge to the Abrikosov vortices and form multi half-valued items.

In the equilibrium the cores of Abrikosov vortices tend to become oriented
perpendicular to the plane of the interface.  Within the non SC layer there arises
cylindrical region with relatively strong magnetic field that reproduces the Abrikosov lattice generated within adjacent SC layers.  Thus, the interlayer mimics a magnetic lattice with localized magnetic momenta.

Let us consider now the case when OAM wave packets travel laterally within such a
non-SC layer with strongly non-uniform magnetization.  Magnetic moment of the OAM
photon (OAM excitation) interacts with magnetic field that penetrates the non-SC layer. Effectively for the travelling  OAM bullets the Abrikosov lattice provides an ordered net of scattering centres.  The magnetic field associated with the Abrikosov influences the motion of OAM wave packets due  to the magnetic interactions. The OAM bullet carries orbital momentum $ \vec{L}_b$ with the "z" component equal $L_b^z=l_b\hbar$, with the Abrikosov vortex there is associated angular momentum $\vec{L}_v$  with the $L_v^z =l_v\hbar$, where $l_v$ and $l_b$ represent the topological charges of the OAM bullet and Abrikosov vortex respectively. With these abbreviations the effective interaction can be written as $H \propto \vec{L}_b \times \vec{L}_v$. This asymmetric interaction creates a novel type of the Hall effect, the {\it OAM Hall effect}, which  deflects asymmetrically the OAM beams and  produces orbital momentum  excess at one of the surfaces. Classical Spin Hall effect being a counterpart of the {\it OAM Hall effect} is composed of three components: skew scattering, side jump and Berry phase contributions. Studies of the OAM light reflection at interface show along with the classical reflection a side shift effect. This indicates that the light {\it OAM Hall effect} should comprise  all of the three components.

The ballistic transport the OAM $\vec{L}$ current $J_o$ classically are given as the
expectation values of the velocity $\vec{v}=i[H,r]_a$ and momentum operators \cite{cheng}
\begin{equation}
\label{e5}
J_o=\frac{1}{2}[v,L_z]_a
\end{equation}
where $[,]_a$ denotes the anticommutator. When we have the strong coupling between magnetic momenta of the AOM packet and those of underlying medium angular momentum currents are not conserved. The effective currents follow the formula $T_\alpha$ is the dipole torque density that varies initial currents. ${\it J}_\alpha =  J_\alpha +T _\alpha$ where $\alpha$ denotes OAM current. The continuity equation for the  effective OAM current requires
\begin{equation}
\label{e8}
\frac{\partial L_z}{\partial t} + \nabla J_o =  T^o_z
\end{equation}
where $ T^o_z $ is the torque density with this the effective OAM current $J_O^{eff}$ reads as
\begin{equation}
\label{e9}
J_o^{eff} = J_o + x\tau _o
\end{equation}
where $x\tau _0$ gives the correction due to the OAM torque.

The Abrikosov vortices are not rigid items thus at the scattering process some portion of momentum and energy of the traveling OAM wave-packet can be transferred  to the vortex and induce its motion. We should remind here that although vortices are commonly assumed to exit the superconductor perpendicular to the surface there were  found tilted vortex lattices in $ NbSe_2$ under action of tilted fields \cite{hess}.
As we are considering a SC/no-SC/SC sandwich  we should note that under action (perpendicular to the surface) of magnetic field or OAM photon a twin vortices in both SC layers are created. Suppose now that in our system with some concentration of  vortices within the non-SC layer travels  an OAM photon or beam. In such a situation, due to the mutual interaction of OAM bullet and Abrikosov vortex there appears asymmetric Hall deflection of the OAM and the vortex shift and/or tilt (see fig. 2).
\begin{figure}[ht]
\includegraphics[height= 4.0cm, width= 6.5cm]{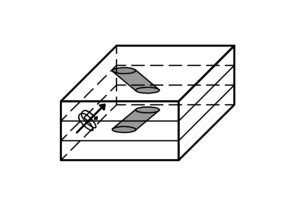}
\caption{Vortex tilt (not in scale) and OAM photon asymmetric deflection due to their mutual interaction \label{e51a}.}
\end{figure}
With the external fields this can even induce a kind of Larmour precession of the vortices. Thus, it is of interest to account the tilt ad possibly induced precession of vortices due to the interaction between the SC vortex and OAM beam since this affects the inter-vortex interactions and Hall effect \cite{kirtley}. The phase of the precession in a row of vortices  differs due to the delay in the time of excitation. In this way there arises a picture in which the precession of vortices mimics the semiclassical spin wave model. However, contrary to the spin waves the precession of vortices is damped and initial orientation of the vortex cores is restored. As we have discussed above the Abrikosov vortices can be created two ways: by the application of external fields or by OAM photons irradiation.  In the field induced vortices the external magnetic field forces tends to restore initial orientation. In both types of Abrikosov vortices there arises dissipation of the energy due to the normal core resistivity. It is well-known that the motion of vortices induces resistance and as a consequence the effective current $J$ swirling the normal core is given by
\begin{equation}
\label{e7q}
J=\left(\frac{1}{\kappa} \nabla \Phi - A\right)|\psi|^2 -\sigma
\left(\frac{\partial A}{\partial t} - \nabla \Phi\right)
\end{equation}
where {\bf A} and $\Phi$ are the vector and scalar potentials while $\sigma$ in the second term denotes the classical conductivity \cite{deang}. The classical component is responsible for the dumping of the vortex precession.

When thinking about Abrikosov vortex we automatically assume the currents swirling around its normal core to have cylindrical symmetry and their monotonic decay with the separation from the centre of vortex. Evidently for the vortices created by the external magnetic field this  is the ground state of phase defect. However, one may expect that the vortices created by the OAM photons may inherit their symmetry. Thus, other form of the Abrikosov vortex with currents reproducing spatial distribution of the LG beams (\ref{e3}) is  worth considering and we can have many excited forms of the vortex. The relaxation to the ground states goes due to the energy loss within the normal core (see Eq. (\ref{e7q})).

\section{Summary and conclusions}

Single OAM photons/excitations are of paramount importance for modern quantum information science \cite{morin}, since the OAM can be used as an additional degree of freedom with a view to achieving high-capacity and high-spectral-efficiency in communication and computational systems. For example OAM wave packets offer efficient control of charge flow by acting on their orbital moment \cite{fert}. Recent progress in microfabrication techniques has brought electronic devices of the order of or smaller than the OAM diffusion length, which provide possibilities of a variety of applications \cite{otani}. Both achievements can be applied to improving existing or generating new technologies and computing concept, widening the potential of solid state orbitronics. This was the main motivation of our research.
 
We have focused on a study of the sandwich composed of a normal material and two superconducting overlayers. The crucial effect is that the vortex light beams/photons can  transfer its orbital angular momentum to the internal dynamics of the solid state system. In this way in the laminar systems the multiquanta Abrikosov vortices can be created. This opens possibility to apply in a computer memories based on Abrikosov lattices. The information can be encoded by multiplying a number of distinguishable states, because an AOM photon can carry an arbitrarily large amount of information distributed over its spin and orbital quantum states and transfer it to the multiquanta Abrikosov vortices.

In our study we have proved that twisted light beams can create multivalued Abrikosov vortices with carrying many fluxons each. Further it was shown that multivalued Abriksov vortices can be serve as an effective media to store information, spin, angular momentum and energy. Moreover in the case of Josephson junction we postulate that half-valued vortices merge to the multivaled ones. Hypothesis that at Josephson junctions multi and semi-valued vortices can be formed needs experimental verification. The same refers to the possibility of collective Larmor precession of the vortices under action of the OAM beams. The respective experimental techniques are in common use. Thus the verification of the hypothesis formed above should be easy.

Another concept bases on the fact that Abrikosov vortex is immersed into a bulk condensate and its supercurrents condensate is not coherent with it. As we know magnetic fields within the vortex favors non s state. Thus electron pairs forming vortex supercurrents can be in different electronic state than this of bulk condensate. Considerations above show that from the basic physics point of view there are still some very important aspects of superconductivity that need to be understood.

\end{document}